\documentstyle[twocolumn,psfig,rotating]{mn}
\def\pmb#1{\setbox0=\hbox{#1}%
\kern-.025em\copy0\kern-\wd0
\kern.05em\copy0\kern-\wd0
\kern-.025em\raise.0433em\box0}

\newcommand{\aap}{Astr. \& Ap.}
\newcommand{\araa}{Ann. Rev. Astr. \& Ap.}
\newcommand{\apj}{ApJ}
\newcommand{\apjl}{ApJL}
\newcommand{\mnras}{MNRAS}
\newcommand{\nat}{Nature}

\newcommand{\be}{\begin{equation}}
\newcommand{\ba}{\begin{eqnarray}}
\newcommand{\ee}{\end{equation}}
\newcommand{\ea}{\end{eqnarray}}
\newcommand{\sgra}{Sgr A*~}

\begin{document}
\title[Feeding Sgr A*]
{Direct Feeding of the Black Hole at the Galactic Center 
with Radial Gas Streams from Close-In Stellar Winds}

\author[Loeb]{Abraham Loeb\\
Astronomy Department, Harvard University, 60
Garden Street, Cambridge, MA 02138, USA\\ E-mail: aloeb@cfa.harvard.edu}

\maketitle

\begin{abstract}
We show that the recently-discovered orbits of massive stars with closest
approach of $\sim 10^3$ Schwarzschild radii from \sgra, allow winds
from these stars to provide the required mass deposition rate near the
black hole horizon. The observed luminosity of \sgra does not require
viscous transport of angular momentum, as long as the total wind mass loss
rate reaches $\sim 10^{-6}M_\odot~{\rm yr^{-1}}$ from the close-in
stars. The specific orbits of the nearest stars should cause modulation of
the radio and infrared flux from \sgra on a timescale of years in a
predictable fashion.
\end{abstract}


\section{Introduction}

Early in the development of accretion flow theory, it had been realized
that angular momentum can severely limit the accretion rate of gas onto a
black hole relative to the spherical infall value (the so-called {\it
Bondi} (1952) {\it rate}).  Indeed, most of the hot gas revealed by recent
X-ray observations within $\sim 0.1$ pc of the Galactic center (Baganoff et
al. 2003), is apparently not flowing into \sgra, the massive black hole
there (Quataert 2003).  The latest models of the {\it radiatively
inefficient accretion flow} around \sgra estimate the accretion rate of gas
into the black hole horizon at $\sim 10^{-8}M_\odot~{\rm yr^{-1}}$, several
orders of magnitude below the Bondi rate (Yuan, Narayan, Quataert 2003).
This low accretion rate is required by the frequency dependence of the
linear polarization data (Bower et al. 2003), in combination with the
synchrotron radio flux and the recently detected infrared emission from
\sgra (Genzel et al 2003b; Ghez et al. 2003c).  The absolute minimum
accretion rate required by the bolometric luminosity of \sgra ($\sim
10^{36}~{\rm erg~s^{-1}}$) is $\sim 2\times 10^{-10}M_\odot~{\rm
yr^{-1}}(\epsilon/0.1)^{-1}$, where $\epsilon$ is the radiative efficiency
of the accreting mass.

The disparity between the large gas reservoir available at the outer radius
of influence of \sgra and the small mass that actually flows into its
horizon, has inspired a variety of accretion models in which the mass
inflow rate diminishes with decreasing radius, due to convection (Quataert
\& Gruzinov 2000a,b; Narayan et al. 2000), turbulence (Coker \& Melia 1997;
Pen et al. 2003), or outflows (Begelman \& Blandford 1999; Yuan et
al. 2002). In this {\it Letter}, we highlight the significance of an
important ingredient that was omitted by these models, namely the powerful
winds from known stars that are embedded within the dilute accretion flow
towards Sgr A*.

\sgra was recently found to be surrounded by a cluster of young, massive
stars.  Over the past year, three of these stars, named SO-2, SO-16, and
SO-19 by Ghez et al. (2003c; see parallel work by Sch{\" o}del et
al. 2003), were inferred to have orbits with peri-centric distances of
$\sim 100$AU or $\sim 10^3$ Schwarzschild radii from the central black
hole. The detection of a Br$\gamma$ and He I features in of the spectrum of
SO-2 indicates that it is probably an O8-B0 star, while the other bright
stars are also likely to be O-B stars, consistent with the lack of a CO
absorption feature in their spectrum (Ghez et al. 2003a; Eisenhauer et
al. 2003).

The winds of O-stars on the main sequence have typical mass loss rates of
${\dot M}_w=10^{-7}$--$10^{-6}M_\odot~{\rm yr^{-1}}$ and speeds in the
range $v_w=1$--$3\times 10^3~{\rm km~s^{-1}}$ (Puls et al. 1996; Repolust,
Puls, \& Herrero 2003).
In \S 2.1, we show that SO-2, SO-16, and SO-19 can supply the required
low-angular momentum gas that fuels the emission from \sgra {\it without
requiring viscous transport of angular momentum}. To demonstrate this
point, we adopt the simplest model in which the radial gas streams move on
ballistic Keplerian orbits. Gas with higher angular momentum is assumed,
for simplicity, to be driven out of the system, as inferred to be case at
the Bondi radius (Quataert 2003).  In \S 2.3 we show that the
characteristic ram-pressure of the radial streams is well above the thermal
pressure of the diffuse hot gas detected by the {\it Chandra X-ray
Observatory} (Baganoff et al. 2003), making the ballistic approach
viable. In \S 2.2 we illustrate how the specific orbital parameters of the
close-in stars can be used to predict the long-term modulation of the radio
and infrared flux of Sgr A*. Finally, we discuss the implications of these
results in \S 3.

\section{Direct Flow of Gas Into the Black Hole Horizon}

\subsection{Method of Calculation}

We consider the simplest model in which the gas elements move on ballistic
Keplerian orbits.  This assumption is justified if: (i) viscosity is
sufficiently weak so that mainly gas on nearly radial orbits approaches the
horizon of \sgra; (ii) the inflow of low angular momentum gas is dominated
by one stellar wind at a time so that multiple gas streams do not collide;
and (iii) the wind ram pressure is much higher than the gas into which it
propagates, so that the low-angular momentum gas is not deflected by
hydrodynamic forces on its radial trajectory towards the black hole
horizon. We justify the latter two conditions in \S 2.3.

The mass of gas directed into the compact emission region around the black
hole horizon is the sum of all gas elements sent out from a stellar wind
with angular momentum per unit mass smaller than some critical value, $J_{\rm
max}$ (see Fig. 1).  We write, 
\begin{equation}
J_{\rm max}= \eta \left({4GM\over c}\right),
\end{equation}
where $M=4\times 10^6M_\odot$ is the black hole mass (Ghez et al. 2003b;
Sch{\" o}del et al. 2003), and $4GM/c$ is the maximum angular momentum with
which the gas would enter directly into the horizon of a Schwarzschild
black hole (see equation 14.2.12 in Shapiro \& Teukolsky 1983). For a
nearly radial trajectory starting at the star, we find that
$\eta=0.5x/(x-1)^{1/2}$, where $x=r_{\rm min}/(2GM/c^2)$ is the distance of
closest approach to \sgra (in Schwarzschild radii units) of the ballistic
gas orbit\footnote{This result is obtained by setting ${\tilde E}\approx 1$
and solving $V(r)={\tilde E}$ in equation 12.4.25 of Shapiro \& Teukolsky
(1983).}.  In popular emission models (Melia et al. 2001; Melia, Liu, \&
Coker 2001; Liu \& Melia 2002a; Yuan et al. 2002; Yuan et al. 2003), most of
the bolometric luminosity of \sgra is produced within $\sim 30$
Schwarzschild radii, implying $\eta\la 2.8$ for the wind gas to join the
emission region directly from its orbit.  Next, let us calculate the
fraction of the wind material that has an orbital angular momentum $\vert
J\vert <J_{\rm max}$ for a star at a radius $r$ in an orbit with an
eccentricity $e$ and a semi-major axis $a$.

\begin{figure}
\centering
\mbox{\psfig{figure=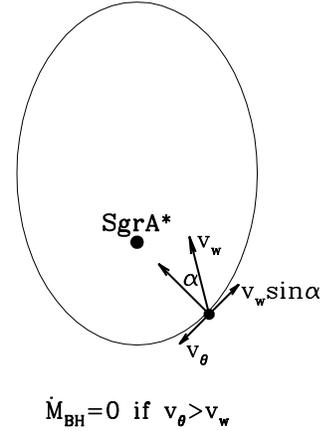,height=3.0in,width=3.0in}}
\caption{Schematic geometry of a radial stream. The stream is confined to a
narrow cone centered on the direction where the tangential component of the
wind velocity, $v_w\sin \alpha$, cancels the tangential velocity of the
star in its orbit around \sgra, $v_\theta$ (see Eq. \ref{eq:ang_mom}). The
small opening angle of the radial stream, $\delta$, is dictated by the
maximum angular momentum, $J_{\rm max}$, that allows the gas to join the
compact emission region around \sgra (Eq. \ref{eq:ang_rad}).}
\end{figure}

The low angular-momentum gas is confined to a narrow cone centered on an
angle $\alpha$ relative to the direction of \sgra (see Fig. 1), such that
\begin{equation}
v_\theta +v_w \sin \alpha= 0,
\label{eq:ang_mom}
\end{equation}
where $v_w$ is the wind speed, $v_\theta=\ell/r$ is the tangential
component of the orbital velocity of the star (using the polar coordinates
$(r,\theta)$ in the orbital plane), and $\ell=[GMa(1-e^2)]^{1/2}$ is the
orbital angular momentum of the star. Note that condition
(\ref{eq:ang_mom}) may be satisfied only through a portion of the stellar
orbit that is sufficiently far from \sgra so that $v_w$ is larger than
$v_\theta\propto r^{-1}$; elsewhere, we set the mass deposition rate into
\sgra to zero (see the gaps of vanishing ${\dot M}_{\rm BH}$ in
Fig. 2). The small opening angle of the \sgra acceptance cone, $\delta
\ll\alpha$, is defined by the condition,
\begin{equation}
v_\theta +v_w \sin(\alpha+\delta)= {J_{\rm max}\over r} .
\label{eq:ang_rad}
\end{equation} 
Using the small-angle approximation, $\sin(\alpha+\delta)\approx \sin\alpha
+\delta\cos\alpha$, we find
\begin{equation}
\delta={J_{\rm max}\over v_w r \cos\alpha}, 
\end{equation}
where $\cos\alpha= \left[1-(\ell/v_w r)^2\right]^{1/2}$ .

We can now evaluate the fraction of the wind mass that is channeled through
the acceptance cone of the emission region around the black hole
horizon. Assuming that the wind outflow is spherically-symmetric in the
star rest-frame, this fraction is
\begin{equation}
{{\dot M}_{\rm BH}\over {\dot M}_w}={\pi \delta^2\over 4\pi}= \left({J_{\rm
max}\over 2 v_w r}\right)^2\left(1-{\ell^2 \over v_w^2r^2}\right)^{-1} .
\label{eq:mdot}
\end{equation}
For the characteristic values of $v_w\sim 1000~{\rm km~s^{-1}}$ and $r\sim
10^3$ AU, one gets ${\dot M}_{\rm BH}\sim 5\times 10^{-3} (\eta/3)^2 {\dot
M}_w$.

If the gas element leaves the star at a time $t$, then it will reach the
black hole on a radial orbit at a time $t+\Delta t$.  The time delay
$\Delta t$ can be easily calculated from the initial radial distance of the
star $r$, and the initial radial velocity of the gas,
\begin{equation}
u_r= {dr\over dt} \pm v_w \left[1-\left({\ell \over v_w
r}\right)^2\right]^{1/2} ,
\label{eq:vradial} 
\end{equation}
where there are either zero, one or two allowed cones under the restriction
that $u_r<0$. The timescale for the gas to be drained into the black hole
horizon from the compact emission region around it (Hawley, Balbus, \&
Stone 2001), is expected to be short compared to the timescales of interest
here and so we neglect it.

\begin{figure} 
\centering
\mbox{\psfig{figure=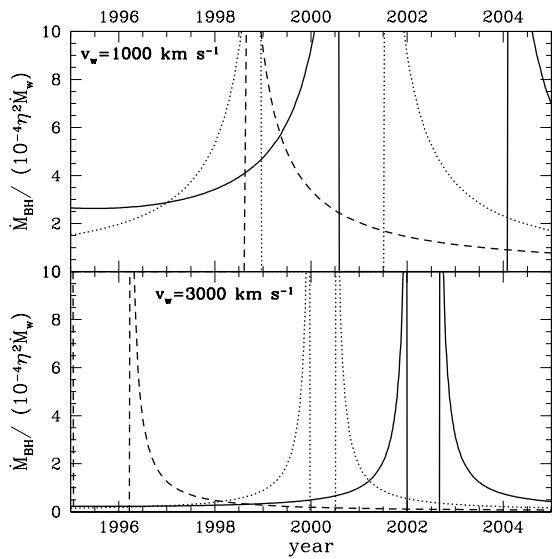,height=3.0in,width=3.0in}}
\caption{Mass deposition rate into the emission region surrounding the
horizon of the black hole at the galactic center, ${\dot M}_{\rm BH}$, as a
function of calender year. Different line types correspond to the
individual contributions of the three bright stars nearest to \sgra, namely
SO-2 (solid), SO-16 (dashed), and SO-19 (dotted).  The orbital parameters
of these stars were adopted from Table 3 of Ghez et al. (2003). The
sensitivity of the results to the assumed wind velocity, $v_w$, is
illustrated by the difference between the {\it top} ($v_w=1000~{\rm
km~s^{-1}}$) and {\it bottom} ($v_w=3000~{\rm km~s^{-1}}$) panels.  The
gaps in ${\dot M}_{\rm BH}$ occur during phases with $v_\theta>v_w$ along
the stellar orbit, and are centered on pericenter passages.  }
\end{figure}

\subsection{Numerical Results}

Figure~2 shows the numerical solution of equations (\ref{eq:mdot}) and
(\ref{eq:vradial}) for ${\dot M}_{\rm BH}$ as a function of $t+\Delta t$,
given the orbital parameters of SO-2 (solid line), SO-16 (dashed), and
SO-19 (dotted), as summarized in Table 3 of Ghez et al. (2003). The
evolution of the mass deposition rate into the black hole horizon is
presented for the values $v_w=1\times 10^3~{\rm km~s^{-1}}$ (upper panel)
and $3\times 10^3~{\rm km~s^{-1}}$ (lower panel), that bracket the range of
wind speeds deduced from observed spectra of O stars (Puls et al. 1996;
Repolust et al. 2003).

For the value of ${\dot M}_w\sim 10^{-6}~M_\odot~{\rm yr^{-1}}$ found in
bright Galactic O stars on the main sequence (Puls et al. 1996; Repolust et
al. 2003), we get a total feeding rate of $\sim
10^{-8}(\eta/3)^2~M_\odot~{\rm yr^{-1}}$, comparable to the value required
by the polarization data in accretion models of \sgra (Yuan et al. 2003;
Melia, Liu \& Coker 2000; Bromley, Melia, \& Liu 2001).  Most of the wind
material spreads over the highly eccentric orbit of its parent star, and so
we find that ${\dot M}_w\la 10^{-6}~M_\odot~{\rm yr}^{-1}$ for SO-2, SO-16,
or SO-19 yields electron densities below the canonical inflow model of Yuan
et al. (2003) with the resultant Faraday rotation measure satisfying
current polarization constraints.

It is natural to expect the luminosity of \sgra to scale with ${\dot
M}_{\rm BH}^\beta$, where $\beta$ ranges between 1 (for disk or jet
emission with a constant radiative efficiency) and 2 (for radiatively
inefficient flows). The emission spectrum can be generically explained by
synchrotron and inverse-Compton processes (Melia \& Falcke 2001; Yuan et
al. 2002; Yuan et al. 2003) in a hot, magnetized gas within tens of
Schwarzschild radii from \sgra, independent of the exact accretion
geometry. The radial streams that reach this region would likely dissipate
a fraction of their kinetic energy through shocks, and settle into a
rotationally supported configuration around Sgr A* (see Hawley, Balbus, \&
Stone 2001, for the subsequent evolution of such as torus). The interaction
of the freely-falling material with the rotating torus may resemble the
situation in wind-fed X-ray binaries (Beloborodev \& Illarionov 2001). The
magnetic fields carried by the infalling streams could be amplified by
plasma instabilities and produce the synchrotron luminosity of Sgr A*.

\subsection{Are Ballistic Orbits Plausible?}

The properties of the accretion flow under consideration here resemble the
conditions in wind-fed X-ray binaries (Illarionov \& Beloborodev 2001,
and references therein).  The accreting matter in such binaries originates
in a wind from an OB star with similar mass loss rates to those assumed
here, and is thought to freely-fall towards the boundary of the accretor
(be it the surface of a neutron star or the horizon of a black hole) where
the X-rays are produced. The ballistic approach adopted here had been used
successfully to derive the accretion rate and the corresponding X-ray
luminosity of accreting neutron stars (Illarionov \& Beloborodev 2001).

Figure~2 illustrates that at most times there is only one gas stream
dominating the mass deposition rate into Sgr A*. Collisions of multiple
streams with comparable momentum flux are rare. The dominant gas streams
follow ballistic orbits as long as their ram pressure, $p_{w}= \rho_w
v_w^2$, exceeds the thermal pressure of the dilute gas through which they
propagate, $p_{\rm th}$. We find,
\begin{equation} 
p_{w} = {{\dot M}_w v_w\over 4\pi d^2}= 3\times 10^3~ {\dot M}_{w,-6}
v_{w,2} \left({ 10^3~{\rm AU}\over d}\right)^{2}~~{{\rm keV}\over {\rm
cm^3}} ,
\end{equation}
where ${\dot M}_{w,-6}\equiv ({{\dot M}_w/10^{-6}M_\odot~{\rm yr^{-1}}})$,
$v_{w,2}\equiv ({v_w/ 2000~{\rm km~s^{-1}}})$, and $d$ is the distance from
the star (ignoring gravitational focusing by \sgra).  The diffuse X-ray
emitting gas, spread over a scale of $\ga 10^4$ AU from \sgra, has a
characteristic temperature of $\sim 1.3$ keV and a density $\sim 30~{\rm
cm^{-3}}$ (Baganoff et al. 2003), implying a background thermal pressure,
$p_{\rm th}\sim 40~{\rm keV~cm^{-3}}$, which is below the ram pressure of
the radially infalling streams at $r\ll 10^4$AU.  As each stream falls
towards the black hole, it gets accelerated and compressed and so its ram
pressure increases. Since its infall speed scales as $\propto r^{-1/2}$ and
its surface area must decrease as $r^{-2}$, we expect that the ram pressure
would scale as $p_w\propto r^{-3}$ and that the condition $p_{w}>p_{\rm
th}$ will be maintained throughout the trajectory of the stream. Our
simplifying assumption (which we intend to critically examine with a direct
numerical simulation in the future) is that most of the outer gas detected
by {\it Chandra} or any other gas with significant angular momentum is not
able to flow inwards similarly to the radial streams in the absence of
efficient angular momentum transport.  Note that a clumpy wind would
enhance the survivability of the radial streams and potentially lead to
short-term variability of Sgr A*.

The fate of wind gas with angular momentum above $J_{\rm max}$ remains an
interesting open question. The linear polarization data (Bower et al. 2003)
implies that most of this gas does not reach the vicinity of the black
hole. Since most of the energy dissipation occurs at $\sim 10$
Schwarzschild radii and the gravitational binding energy per particle
decreases considerably at larger radii, we conjecture that heat flux from
the inner regions unbinds the outer gas and expels it in an outflow
(similarly to the conditions at the Bondi radius; see Quataert 2003).

\section{Discussion}

We have found that radial gas streams from the winds of massive stars with
recently discovered orbits that approach within $\sim 10^3$ Schwarzschild
radii of the black hole at the Galactic center, can supply the required
mass flow near the central black hole horizon and power the radio and
infrared emission of Sgr A*. For O star winds with a total ${\dot M}_w\sim
10^{-6}M_\odot~{\rm yr^{-1}}$, we get that SO-2, SO-16, and SO-19 can
supply as much as $\sim 10^{-8}(\eta/3)^2 M_\odot~{\rm yr^{-1}}$ of gas in
orbits directed straight into the emission region around the black hole
horizon. The total mass inflow rate would increase if angular momentum of
other gas elements is transported away by magnetic or turbulent stresses.
The surprising existence of massive stars in orbits with closest approach
distances of $\sim 100$ AU poses a major puzzle to theories of star
formation (Genzel et al. 2003a), and was not incorporated in previous
modeling of the gas flow around \sgra (e.g. Coker \& Melia 1997; Yuan
et al. 2003).

The parameters of the winds could in principle be determined through
detailed spectroscopic observations of the stars in the \sgra cluster. Once
the wind properties of the individual stars are constrained, our accretion
model will be easily testable since it correlates the long-term ($>$month)
variability of the radio or infrared emission from \sgra with the orbital
phases of the stars around it (see Fig. 1).  The simplified calculation
presented in this {\it Letter} can be improved using a full hydrodynamic
simulation and compared against observational data on the long-term
variability of \sgra (Zhao et al. 2003) for different choices of $v_w$ and
${\dot M}_w$ per individual star (for an alternative source of this
variability, see Liu \& Melia 2002b). A proper numerical treatment is
challenging since only a small fraction of the gas mass follows the radial
orbits of interest here, and since the inflow spans several orders of
magnitude in radius - requiring a large dynamic range in spatial and
temporal resolution. Previous accretion simulations (such as those by
Coker, Melia, \& Falcke 1997, Proga \& Begelman 2003 or Rockefeller et
al. 2003) did not meet the demanding requirements of this accretion
problem.

\sgra may be surrounded by fainter stars that have not been detected so
far, even though they dominate the steady mass flow into the black
hole. However, winds from massive stars are driven by radiation pressure,
and so the stars dominating the fluctuations in the mass flow into the
black hole horizon should have the highest luminosities and be included in
the current census of bright infrared sources near Sgr A*.  Massive stars
that are much further away also possess powerful winds (Krabbe et al. 1991;
Najarro et al. 1997; Coker \& Melia 1997; Quataert 2003), but their
individual contributions to the radial mass flux near the horizon of \sgra
is reduced compared to the close-in stars\footnote{Statistically, the net
mass flux is dominated by the close-in stars as long as the number density
run of wind sources is steeper than $\propto r^{-1}$, as is the case around
\sgra (Genzel et al. 2003a).}, since ${\dot M}_{\rm BH}\propto r^{-2}$ in
equation~(\ref{eq:mdot}). For example, IRS 16C with its ${\dot M}_w\approx
8\times 10^{-5} M_\odot~{\rm yr^{-1}}$ and $v_w \sim 650~{\rm km~s^{-1}}$
at $r\ga 2\times 10^4$AU, and IRS 13E1 with its ${\dot M}_w\approx 8\times
10^{-4} M_\odot~{\rm yr^{-1}}$ and $v_w \sim 1000~{\rm km~s^{-1}}$ at $r\ga
4\times 10^4$AU (Quataert, Narayan, \& Reid 1999), produce at most a radial
mass flux comparable to that of SO-2 if the latter has ${\dot M}_w\sim
10^{-6}M_\odot~{\rm yr^{-1}}$.  Nevertheless, the steady mass flux into
\sgra may still be dominated by undetected stars or by gas that originates
at large distances and loses its angular momentum through viscous transport
(Yuan et al. 2003). In this case, the winds from the close-in stars would
only modulate a fraction of ${\dot M}_{\rm BH}$, but their fractional
contribution could still be identified through the predictable short-term
variability that their orbits introduce to the luminosity of Sgr A*.

\bigskip
\noindent
{\bf Acknowledgements}

I thank Jean-Pierre Lasota, Eric Pfahl, Eliot Quataert, Mark Reid, George
Rybicki, and Eli Waxman for comments on the manuscript.  This work was
supported in part by NASA grant NAG 5-13292, and by NSF grants AST-0071019,
AST-0204514.

\vfil\eject
\newpage
\newpage

\end{document}